\documentclass[epjCONF]{svjour}
\usepackage{graphics}
\usepackage[varg]{txfonts} % Times fonts
\usepackage[latin1]{inputenc}
\session-title{Search for the SM Higgs decaying to a \emph{b}-quark pair with ATLAS}
\begin{document}
\title{Search for the Standard Model Higgs boson decaying to a \emph{b}-quark pair with ATLAS}
\author{Alberto Palma \inst{1,}\inst{2}\fnmsep\thanks{\email{apalma@lip.pt}}, on behalf of the ATLAS Collaboration}
\institute{FCUL -- Faculdade de Ci\^encias da Universidade de Lisboa \and LIP -- Laborat\'orio de Instrumenta\c c\~ao e F\'isica Experimental de Part\'iculas de Lisboa}
\abstract{Summary of first ATLAS results is given for the Standard Model Higgs boson in the mass range $110 < m_{H} < 130$~GeV,  when produced in association with a $W/Z$ boson and decaying to a pair of $b$-quarks. No evidence for the Higgs boson production is observed in a dataset of proton-proton collisions at a center of mass energy of 7 TeV and corresponding to 1.04 fb$^{-1}$ of integrated luminosity, recorded by the ATLAS experiment at the LHC in 2011. Upper limits on Higgs boson production cross sections for the channels considered are presented.} %end of abstract
\maketitle
\section{Introduction}
\label{intro}

The search for the Standard Model (SM) Higgs boson is one of the primary objective of the ATLAS  experiment ~\cite{RefATLAS} at the Large Hadron Collider (LHC). If $m_H<2\times m_W$, the $H\rightarrow b\bar{b}$ decay is predominant~\cite{RefXsec}. In this mass region, the dominant production process is gluon fusion ($gg\rightarrow H$), but this channel is challenging to observe due to the enormous QCD multijet background. As opposed to that, the associated production of the Higgs boson with a $W/Z$ boson, despite of its smaller cross section might be observable due to the clear signatures from the leptonic decays of the vector bosons. The study reported here employs a simple and robust cut-based analysis to search for the Higgs boson in the $WH\rightarrow l\nu b\bar{b}$ and $ZH\rightarrow llb\bar{b}$ channels, where the lepton $l$ is an electron $e$ or a muon $\mu$~\cite{RefNota}. These channels are characterized by the presence of one or two isolated leptons with high transverse momenta ($p_{T}$) and two $b$-jets in the final state. In the case of the $WH$, there is also a large amount of missing transverse energy ($E^{miss}_{T}$ ), due to the escaping neutrino $\nu$. The most important backgrounds are due to $W/Z+jets$, QCD multijet, top quark ($t \bar{t}$ or single top) and diboson production ($WW, WZ, ZZ$). Although the cross section of the $ZH$ channel is about half of the $WH$, it is less affected by the top background, and therefore both channels can reach a similar sensitivity. 

The data sample used was recorded by the ATLAS experiment during the 2011 LHC run at a centre of mass energy of $\sqrt{s} = 7$~TeV and represents an integrated luminosity of $\mathcal{L}=1.04$~fb$^{-1}$. 

\section{Event selection}
\label{sec:1}

Events were selected using a single lepton trigger with a $p_T$ threshold of 18~GeV for muons and 20~GeV for electrons. To increase efficiency in the electron channel for the $ZH$ analysis, the single lepton trigger is complemented with a $p_T > 12$~GeV di-electron trigger. Only events with a primary vertex with at least three tracks are selected for further analysis.

Electron candidates are reconstructed from electromagnetic calorimeter clusters matched to tracks reconstructed in the inner tracking detector (ID). Muon candidates are reconstructed by matching tracks found in the ID with either tracks or hit segments in the muon spectrometer (MS). In order to suppress leptons produced in jets, the sum of track transverse momenta in an $\eta-\phi$ cone of radius 0.2 around the lepton track must be smaller than 10\% of the lepton $p_T$. To further reduce semileptonic decays in the $W$ channel, the transverse (longitudinal) distance of the lepton track to the event vertex must be less than 0.1~(10)~mm and muons are required to have an $\eta-\phi$ distance greater than 0.4 to any jet with $p_T>25$~GeV.

Jets are reconstructed from topological energy clusters in the calorimeter using an anti-$k_T$ algorithm, with a radius parameter $R = 0.4$, and are calibrated to the hadronic energy scale. In order to reduce the sensitivity to pile-up from additional $pp$ interactions occurring in the same or out-of-time bunch crossings, only jets associated to the main interaction vertex are used. ATLAS $b$-tagging algorithms are used to identify jets containing decays of $b$-hadrons. In this analysis a combination of the three dimensional impact parameter information and the output of a secondary vertex finding algorithm is used. The $b$-tagging cut is chosen so that an efficiency of 70\% for $b$-jets in Monte Carlo (MC) simulated $t\bar{t}$ events is obtained, while providing a light jet rejection factor of around 50.

The $E^{miss}_{T}$ is measured from the vector sum over all topological clusters in the calorimeters with $|\eta| < 4.5$, together with terms accounting explicitly for selected muons in the event.

To continue the selection in the $ZH\rightarrow llb\bar{b}$ search channel it is required that there must be exactly two leptons in the event that form a $Z$ boson candidate, with invariant mass $76 < m_{ll} < 106$~GeV and $E^{miss}_{T} < 50$~GeV. At least two jets with $p_T > 25$~GeV are required, where the two highest-$p_T$ jets are both required to pass the $b$-tagging selection. 

$WH\rightarrow\nu lb\bar{b}$ events are required to have exactly one lepton. The $E^{miss}_{T}$ in the event is required to be greater than 25~GeV and the transverse mass\footnote{The transverse mass is defined as $m_T =\sqrt{2p^{l}_{T}p^{\nu}_{T}(1-\cos(\phi^{l}-\phi^{\nu}))}$.} should be greater than 40~GeV. The number of jets with $p_T > 25$~GeV is required to be exactly two, to reduce background from top production that is significantly higher in this channel. In addition the two jets must pass the $b$-tagging selection. 

\begin{figure}
\centering
\resizebox{0.43\columnwidth}{!}{%
  \includegraphics{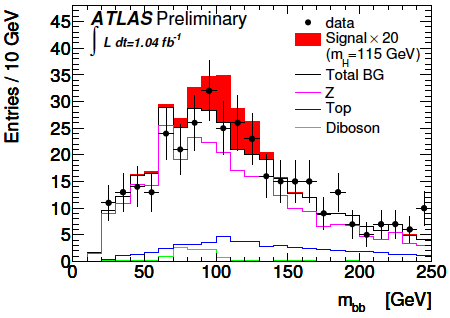} }~\hspace{5mm}~
\resizebox{0.43\columnwidth}{!}{%  
  \includegraphics{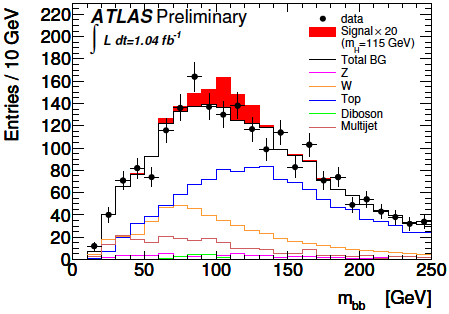} }
\caption{The di-$b$-jet invariant mass for $ZH\rightarrow llb\bar{b}$ (left) and $WH\rightarrow l\nu b\bar{b}$ (right) for $m_H = 115$~GeV. The signal distribution is enhanced by a factor of 20 for visibility~\cite{RefNota}.}
\label{fig:1}       % Give a unique label
\end{figure}

The $m_{b\bar{b}}$ distribution for events surviving the $WH\rightarrow l\nu b\bar{b}$ and $ZH\rightarrow llb\bar{b}$ selections is shown in Figure~\ref{fig:1} for real and MC events. Where possible, control regions are used to determine or verify the normalization and shape of different backgrounds in the data.

\section{Systematic uncertainties}
\label{sec:2}

%The main detector-related contributions to the systematic uncertainties are from the lepton identification efficiency, their energy or momentum scale and resolution, as well as from the jet energy scale and resolution, $b$-tagging efficiency and mis-tagging rates and the trigger efficiencies. The uncertainty on the Higgs production cross section was estimated to be 5\% for both $WH$ and $ZH$ production processes. The uncertainty on the normalization of the $Z+jets$ background for both analyses is 9\% determined by the statistical uncertainty in the control data sample for this background in the $ZH$ analysis. Events in the control sample are required to pass the common selection, but lie in sidebands of the di-lepton mass distribution, defined as $60<m_{ll}<76$~GeV or $106<m_{ll}<150$~GeV. Similarly, a normalization uncertainty of 14\% is assigned to the $W+jets$ background based on the statistical precision of the normalization measurement in the corresponding control data region. In both cases, additional terms to account for the uncertainty on the shape of the $W/Z+jets$ distributions are computed. The normalization uncertainty for the QCD multijet background is taken to be 100\% for $ZH$ and 50\% for $WH$. The normalization error for the top background is 9\% for the $ZH$ and 6\% for the $WH$ analysis, based on the measurements in the corresponding control regions. 

The main detector-related contributions to the systematic uncertainties are from the lepton identification efficiency, their energy or momentum scale and resolution, as well as from the jet energy scale and resolution, $b$-tagging efficiency and mis-tagging rates and the trigger efficiencies. The uncertainty on the Higgs boson production cross section was estimated to be 5\% for both $WH$ and $ZH$ production processes. The uncertainty on the normalization of the $Z+jets$ background for both analyses is 9\% determined by the statistical uncertainty in the control data sample for this background in the $ZH$ analysis. Similarly, a normalization uncertainty of 14\% is assigned to the $W+jets$ background based on the statistical precision of the normalization measurement in the corresponding control data region. In both cases, additional terms to account for the uncertainty on the shape of the $W/Z+jets$ distributions are computed. The normalization uncertainty for the QCD multijet background is taken to be 100\% for $ZH$ and 50\% for $WH$. The normalization error for the top background is 9\% for the $ZH$ and 6\% for the $WH$ analysis, based on the measurements in the corresponding control regions. 

The uncertainty in the integrated luminosity has been estimated to be 3.7\%. This uncertainty is only applied to MC samples for which the normalization error is not taken directly from a comparison between data and MC. 

The most important source of systematic uncertainty is the $b$-tagging efficiency, which affects the signal yields of both channels by 17\%.

\section{Results}
\label{sec:3}

\begin{figure}
\centering
\resizebox{0.4\columnwidth}{!}{%  
  \includegraphics{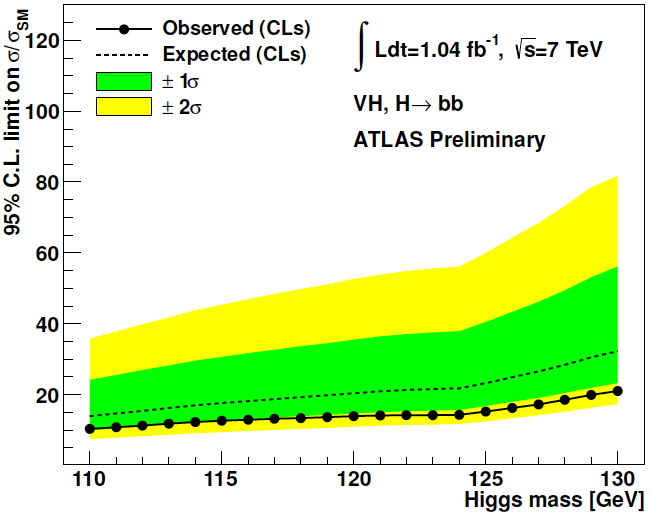} }
\caption{Expected (dashed) and observed (solid line) exclusion limits for the $ZH\rightarrow llb\bar{b}$ and $WH\rightarrow l\nu b\bar{b}$ channels combined, expressed as the ratio to the SM cross section, using the profile-likelihood method with $CL_s$~\cite{RefNota}. The green and yellow areas represent the 1s and 2s ranges of the expectation in the absence of a signal.}
\label{fig:2}       % Give a unique label
\end{figure}

The $ZH\rightarrow llb\bar{b}$ and $WH\rightarrow\nu lb\bar{b}$ analyses are performed for five Higgs boson masses in the range $110<m_H<130$~GeV. The Higgs boson signal is searched in the $m_{b\bar{b}}$ distribution (Figure~\ref{fig:1}). For each Higgs boson mass, a one-sided upper-limit is set on the cross section ratio $\mu = \sigma/\sigma_{SM}$ at a 95\% confidence level (C.L.), where $\sigma_{SM}$ is the cross section predicted by the Standard Model. The exclusion limits are derived from the $CL_s$~\cite{RefCL} treatment of the $p$-values computed with the profile likelihood ratio, using the binned $m_{b\bar{b}}$ distribution for the evaluation of the test statistics. The systematic uncertainties are treated as nuisance parameters and shape uncertainties are treated via morphing. The combined exclusion limit for both channels, shown in Figure~\ref{fig:2}, ranges between 10 and 20 times the standard model cross section, depending on the $m_H$. 

\section{Conclusions}
\label{sec:4}

The ATLAS collaboration presents first results of the direct search for the SM Higgs boson decaying to $b\bar{b}$. No evidence of the Higgs boson was found in a $pp$ collision data sample of $\mathcal{L}=1.04$~fb$^{-1}$ at $\sqrt{s} =7$~TeV. Instead, upper limits on the Higgs boson production cross section of between 10 and 20 times the SM value were determined, in a mass range $110<m_H<130$~GeV.

\end{document}